\documentclass[english,12pt]{article}
\usepackage[T1]{fontenc}
\usepackage[latin1]{inputenc}
\usepackage{times}

\newcommand{\inst}[1]{$^{#1}$}
\usepackage{graphics}
\begin{document}
\title{Dilepton production in $pp$ and $CC$ collisions with HADES}
\date{}
\maketitle

I.Fr\"{o}hlich\inst{10}, 
G.Agakishiev\inst{11}, C.Agodi\inst{1}, A.Balanda\inst{5}, G.Bellia\inst{1,2}, D.Belver\inst{19}, 
A.Belyaev\inst{9}, A.Blanco\inst{3}, M.B\"{o}hmer\inst{15}, J.L.Boyard\inst{17}
% ...
\\
$^{1}$) Istituto Nazionale di Fisica Nucleare - Laboratori Nazionali 
del Sud, 95125 Catania, Italy\\
$^{2}$) Dipartimento di Fisica e Astronomia, Universit\`{a} di Catania, 
95125, Catania, Italy\\
$^{3}$) LIP-Laborat\'{o}rio de Instrumenta\c{c}\~{a}o e F\'{\i}sica Experimental 
de Part\'{\i}culas, Departamento de F\'{\i}sica da Universidade de 
Coimbra, 3004-516 Coimbra, PORTUGAL.\\
% ...
$^{5}$) Smoluchowski Institute of Physics, Jagiellonian University 
of Cracow, 30059 Cracow, Poland\\
% ...
$^{9}$) Joint Institute of Nuclear Research, 141980 Dubna, Russia\\
$^{10}$) Institut f\"{u}r Kernphysik, Johann Wolfgang Goethe-Universit\"{a}t, 
60486 Frankfurt, Germany\\
$^{11}$) II.Physikalisches Institut, Justus Liebig Universit\"{a}t 
Giessen, 35392 Giessen, Germany\\
% ...
$^{15}$) Physik Department E12, Technische Universit\"{a}t M\"{u}nchen, 
85748 Garching, Germany\\
% ...
$^{17}$) Institut de Physique Nucl\'{e}aire d'Orsay, CNRS/IN2P3, 91406 
Orsay Cedex, France\\
% ...
$^{19}$) Departamento de F\'{\i}sica de Part\'{\i}culas. University of 
Santiago de Compostela. 15782 Santiago de Compostela, Spain\\
% ...

\begin{abstract}
{ Dilepton production has been measured with HADES, the ``High
  Acceptance DiElectron Spectrometer''. In $pp$ collisions at 2.2GeV kinetic
  beam energy, exclusive $\eta$ production and the Dalitz decay $\eta\to
  \gamma e^{+}e^{-}$ has been reconstructed. The electromagnetic form factor
  is well in agreement with existing data. In addition, 
  an inclusive $e^+e^-$ spectrum from the
  $^{12}$C+$^{12}$C reaction at 2AGeV is presented
  and compared with a thermal model.}
\end{abstract}

\section{Introduction}
\label{intro}
One of the main open questions in QCD is the origin of the hadron masses.
Beside the so-called Goldstone bosons like $\pi$ and $\eta$, the typical mass
scales of hadrons in the vacuum is in the order of GeV, whereas the current
quark masses $m_u$, $m_d$ are 5-15MeV.  Based on this fundamental question it
has been proposed that the mass of the hadrons is related to the spontanous
breaking of chiral symmetry, which should be partially restored at finite
baryon and energy densities.  In connection to this question, one of the
topics in the modern hadron physics is how hadrons behave in strong
interacting medium.  This means that their properties - like mass and width -
have to be measured inside either cold, dense or hot nuclear matter. In this
context, vector mesons have been proposed as an ideal probe for such studies,
since they decay via an intermediate virtual photon $\gamma^*$ into dileptons
($e^+e^-$ or $\mu^+\mu^-$) which do not undergo strong interaction.  Recently,
the NA60 collaboration~\cite{na60} has extracted the $\rho-\omega$ line shape
using the dimuon channel in In+In collisions at 158AGeV, which would
correspond to a hot environment. These measurement indicates broadening of the
line shape rather than a dropping of the mass.

However, for a complete understanding of the hadronic properties it is
important to measure not only the hot, but also the dense region of the phase
space. At beam energies of 1-2 AGeV, which corresponds to moderate densities
(2-3 $\rho_0$), the production of mesons is dominated by multi-step exitations
of a limited number resonances and their subsequent decays, like
$\Delta^{+,0}\to N\pi^{0}\to N\gamma e^{+}e^{-}$ ($\pi$-Dalitz), $\Delta\to
Ne^{+}e^{-}$ ($\Delta$-Dalitz), $N^{*}(1535)\to N\eta\to N\gamma e^{+}e^{-}$
($\eta$-Dalitz) and the decay of virtual resonances in $N(\omega,\rho)$.
Here, most of the production mechanisms are at or even below threshold, which
means that mesons are more likely produced in the dense phase of the fireball
evolution.

The result of such experiments is usually a dilepton invariant mass spectrum
containing all these sources (di\-lepton-cocktail). Before a conclusion on the
properties of $\rho$ and $\omega$ can be drawn, the contribution of Dalitz decays have to be
subtracted.

\section{Properties of the virtual photon}
\label{prop}

For a more detailed view, it is very helpful to know that a virtual photon
(decaying into 2 stable particles) has 6 degrees of freedom: First, its
invariant mass $M^{inv}_{\gamma^*}$. Moreover the momentum $P^{X}_{\gamma^*}$,
the polar $\theta^{X}_{\gamma^*}$ and the azimuthal angle $\phi^{X}_{\gamma^*}$ of the
virtual photon in the rest frame of the source $X$. Angles are defined with
respect to the production plane and momentum transfer.  Finally, the 2 decay
angles of the photon into the dilepton pair, which are usually described with
the helicity angle $\theta^{ee}_e$ (which is the angle of one lepton in the
dilepton rest frame, with respect to the direction of the photon), and the
Treiman-Yang angle $\phi^{ee}_e$ (the orientation of the decay plane around
the photon direction). Since a detector has a finite acceptance as a
function of total phase space, for an interpretation of the invariant mass
spectrum $M^{inv}_{\gamma^*}$ the additional distribution functions have to be
known.  For the pseudoscalar mesons, these are well under control: For a spin-less
state, no alignment information can be carried from the production mechanism
to the decay, so $\theta^{X}_{\gamma^*}$, $\phi^{X}_{\gamma^*}$ and
$\phi^{ee}_e$ are isotropic. The helicity angle is proposed to be
$1+\cos^{2}\theta^{ee}_e$~\cite{brat}.  For a given mass of $\gamma^{*}$ its
momentum is fixed by the mass of the meson. The only degree of freedom 
is the mass spectrum, which is based on a well-know electromagnetic
form factor~\cite{landsberg}. In total, the pseudoscalar mesons have no
uncertainty and are a good choice for a detector performance test.

This is very different for the $\Delta$ Dalitz decay, because for particles
carrying spin, production and decay do not factorize. In addition, the form
factor as well as the branching ratio are based on calculations~\cite{ernst}
and the helicity angle has quite some uncertainties~\cite{brat}. For the
direct decay of vector mesons, the helicity angle has to be isotropic, but the
Treiman-Yang angle could contain higher order contributions. Moreover, the
$\omega\to\pi^0e^+e^-$ transition form factor cannot be consistently described
within the vector meson dominance model.  This shows, that it is not only
important to measure the dilepton production in heavy ion collision, but
collect also information of the individual sources to avoid systematic errors
based on the detector acceptance. This is one of the goals of the HADES
detector (GSI, Darmstadt).  Therefore, its program spans from $pp$, $\pi p$ to
$\pi A$ and $AA$ collisions. First data has been taken in $CC$ collisions at 1 and
2AGeV, $ArK+Cl$ collisions at 1.78AGeV, and $pp$ at 1.25 and 2.2GeV.

A more detailed view of the detector can be found in~\cite{schicker,nim}.
Shortly, HADES is a magnetic spectrometer, consisting of up to 4 mini drift
chambers (MDC) with an toroidal magnetic field. Particle identification is done
based on time-of-flight measurements. In addition, a Ring Imaging Cherenkov
detector (RICH) and an electromagnetic Pre-Shower detector adds lepton
identification features.

\section{The $pp\to pp \eta$ reaction at 2.2GeV}
\label{eta_pp}

In January 2004 the first $pp$ run has been carried out at a kinetic beam
energy of 2.2GeV, in order to study the
performance of the detector using the exclusive reaction $pp\to pp \eta$ which
has been measured by the DISTO collaboration before~\cite{disto} at kinetic
beam energies of 2.15, 2.5 and 2.85GeV.  Since the decays $\eta\to
\pi^+\pi^-\pi^0$~\cite{cbarrel} and $\eta\to \gamma
e^{+}e^{-}$~\cite{landsberg} are known, it served as a calibration reaction
and first test if HADES is able to measure electromagnetic form factors. The
analysis techniques are described in detail elsewhere~\cite{spataro}.
Basically, for both reaction types kinematical contraints were used to
identify the missing particle $\pi^0$ or $\gamma$, respectively. In addition,
a kinematic refit reduced the background.

\subsection{The hadronic decay $\eta\to\pi^+\pi^-\pi^0$}
\label{hadr_eta_pp}

\begin{figure}
\begin{center}
\resizebox{\columnwidth}{!}{%
  \includegraphics{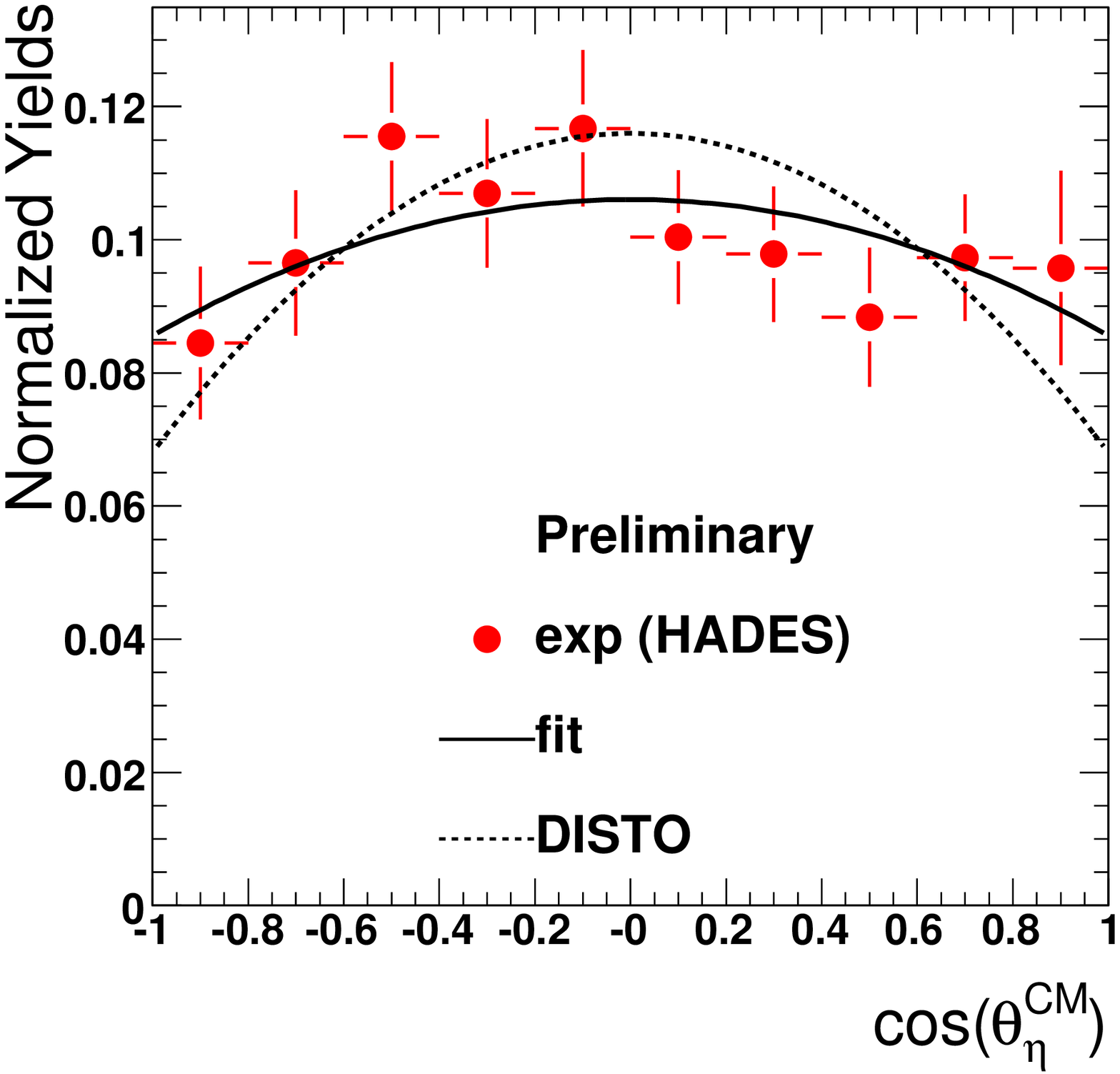}
}
\caption{Distribution of the $\eta$ polar production angle in the center of
  mass frame (statistical errors only). The plot is corrected for acceptance.
  The solid line represents a fit to the data, whereas the dashed lines shows
  a parametrization for the existing~\cite{disto} in the same beam energy
  regime.}
\label{fig:1}       
\end{center}
\end{figure}

By using a selection on $\cos(\theta^{CM}_\eta)$ and fitting the corresponding
$pp$ missing-mass spectrum, the angular distribution of the $\eta$ meson
emission has been evaluated.  The acceptance correction has been done using a
full Monte-Carlo simulation with generated events of $pp\to pM\to pp\eta$
where the shape $M\to p\eta$ was taken from~\cite{disto}. Only the alignment
of the $pp\eta$ plane has been left according to phase space.  Together with
the fact that the beam axis has rotational symmetry, only two angles are left
for the alignment of the $pp\eta$ plane, which are almost independent, thus
allowing to compare directly 1-dimensional angular distributions.  After the
full analysis chain, generated and
accepted events have been divided as a function of $\cos(\theta^{CM}_\eta)$.
Results are shown in Fig.~\ref{fig:1} together with the found distribution by
DISTO with an anisotropy coefficient (2$^{nd}$-order Legendre polynomial) of
$c_2=-0.32\pm0.10$. A fit on the data gives $c_2=-0.14 \pm 0.09$, which tends
more to an isotropic $\eta$ production, but is still consistent within error
bars with the previous result, showing that the hadron efficiency is
understood as a function of phase space.

\subsection{The Dalitz decay $\eta\to \gamma e^{+}e^{-}$}
\label{dal_eta_pp}

\begin{figure}
\begin{center}
\resizebox{\columnwidth}{!}{%
  \includegraphics{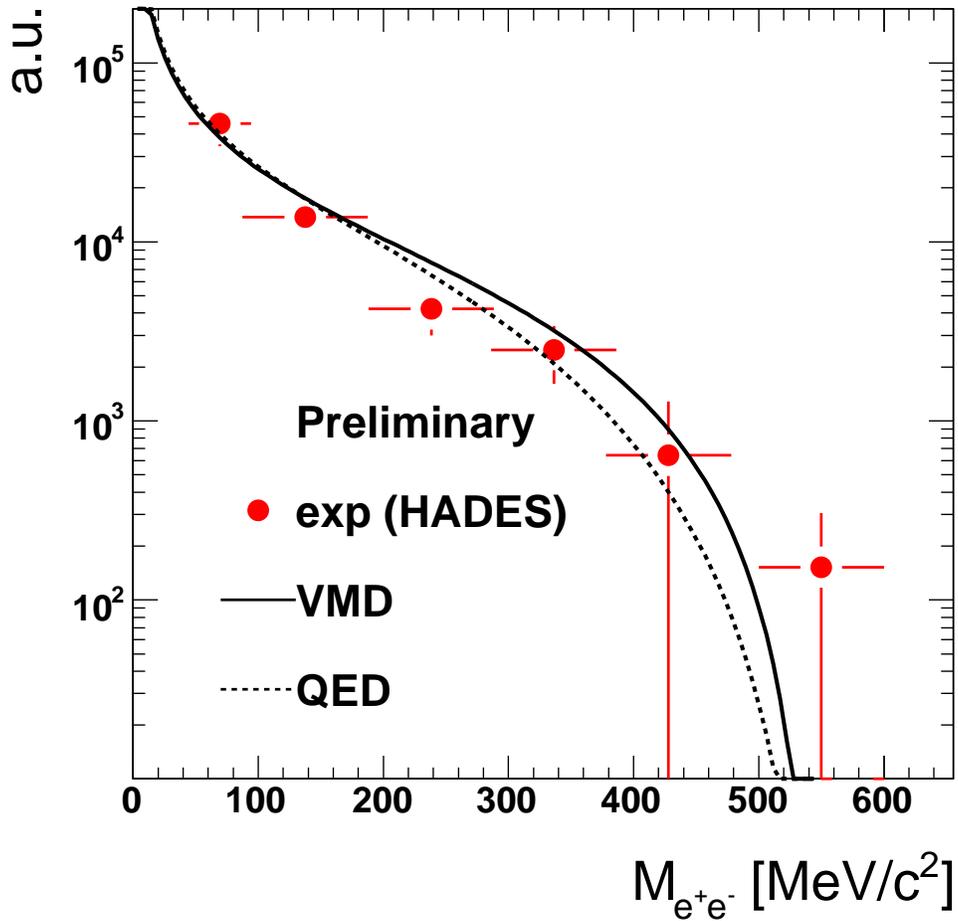}
}
\caption{Invariant mass distribution $M^{inv}_{ee}$ for the decay $\eta\to
  \gamma e^{+}e^{-}$~\cite{spruck} (statistical errors only).
  The dashed line is showing the prediction for a simple QED factor, while the
  solid one represents the full VDM calculation~\cite{landsberg}.
}
\label{fig:2}       
\end{center}
\end{figure}

Similar methods have been used for the $\eta$ Dalitz decay. In addition, a
selection on the opening angle of larger 4$^o$ was used to reject
contributions by $\eta\to\gamma\gamma$ and a subsequent conversion in the
detector material.  The invariant mass was used as the observable which was
corrected for acceptance and efficiency.  Again, by fitting the $pp$
missing-mass spectrum for each invariant mass slice, the yield has been
extracted. For the acceptance correction, the known production angles have
been fixed in the simulation, and the known $\eta$ decay properties have been
used as described in Sec.~\ref{prop}.  Fig.~\ref{fig:2} shows the corrected
invariant mass spectrum~\cite{spruck} together with functions using the QED as
well as vector meson dominance model (VDM) form factor~\cite{landsberg} which
have been normalized to the number of $\eta$'s found by experiment.  In
addition, each data point has been corrected in position according to the
strong slope of the models.  It can be seen that HADES is not sensitive to
distinguish these 2 models, but at least agrees to the predictions within the
given errors. This demonstrates that the efficiency is understood as a
function of the invariant mass, which is the main important observable in the
heavy ion data. In addition, the ratio between the 2 $\eta$ decay channels
(R=$\frac{N_{\eta\to\pi^+\pi^-\pi^0}}{N_{\eta\to \gamma e^{+}e^{-}}}$) have
been calculated for a full-cocktail simulation and the data. The result is
$R_{exp}=15.3\pm1.8_{stat}$ and $R_{sim}=15.6\pm0.9_{stat}$, a nice agreement
confirming that the lepton response of the HADES spectrometer is well
understood~\cite{spataro}.

\section{Results on C+C at 2AGeV}
\label{cc}

The first result on dilepton production in C+C collisions at a kinetic beam
energy of 2AGeV is ready for publication~\cite{prl}. 
Details of the basic analysis steps
can also be found in~\cite{eberl}. Basically, leptons are identified via the RICH
detector and reconstructed using the MDCs. An opening angle of the pair of
9$^o$ has been applied to remove background from conversion. Finally, the
combinatorial background (i.e. pairs mixed from different sources) has been
removed by the like-sign method. For events with masses ($M^{inv}_{ee} > 0.5 GeV/c^2$)
the combinatorial background has been evaluated using the event-mixing
technique.

\begin{figure}
\begin{center}
\resizebox{\columnwidth}{!}{%
  \includegraphics{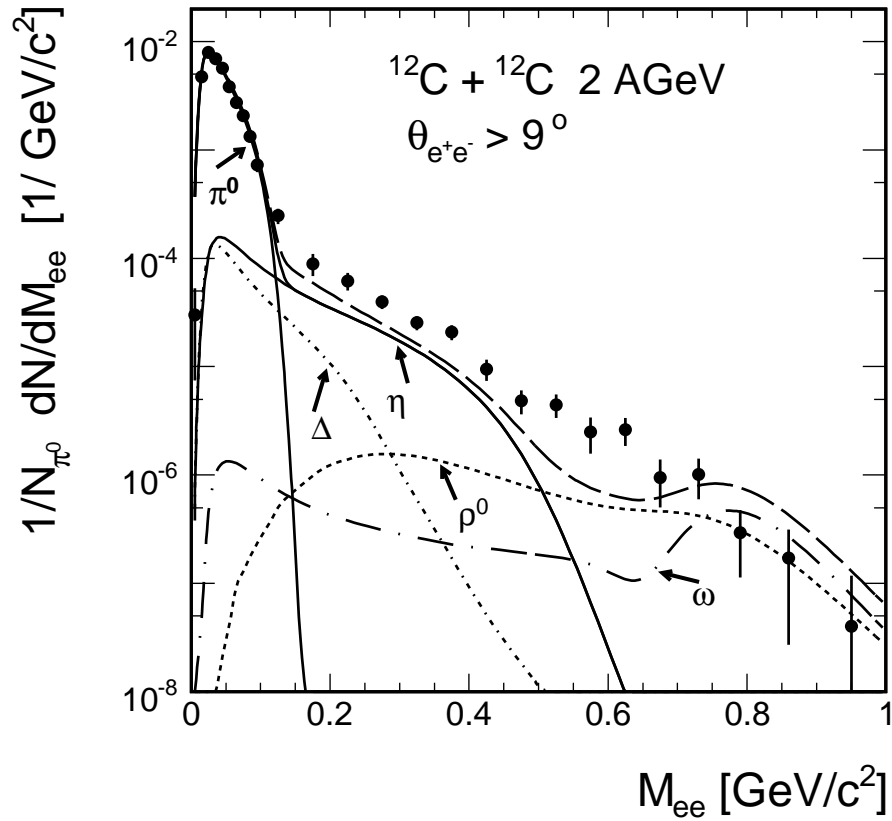}
}
\caption{Invariant mass distribution $M^{inv}_{ee}$ for $CC$ at 2AGeV~\cite{eberl}. 
  The data points contain statistical error only.  The lines are a thermal
  model simulation as described in the text}
\label{fig:3}       
\end{center}
\end{figure}

In contrast to exclusive reactions, in heavy ion reactions only the final
dilepton cocktail can be shown, without any selection on the different
sources. As explained above, the virtual photon properties might vary, thus
the extrapolation to $4\pi$ is difficult, since 6 independent variables should
be taken into account. While the integration over these variables cannot be
done, a different method has been chosen.  First, a single track efficiency
as a function of $p^{lab}_e,\theta^{lab}_e,\phi^{lab}_e$ has been extracted by
simulation and applied to the data. This reduces the acceptance to a binary
value. In order to compare the efficiency corrected data to any model, an
acceptance filter has been developed~\cite{haft}.  
Fig.~\ref{fig:3} shows the
invariant mass spectrum after all analysis steps. Absolute normalization has
been done by a measurement of the charged pions, whose mean is the number of
$\pi^0$ in a isospin-symmetic reaction. 

In order to make a
statement for the contribution of the vector mesons, a simulation has been
made based on a simple thermal model~\cite{pluto} and filtered with the
detector acceptance. The resulting curves are shown also in Fig.~\ref{fig:3}.
The solid line contains the $\pi$ and $\eta$ Dalitz decays with the decay
properties as described above and the known production cross sections and
distribution from TAPS~\cite{taps}. It can be seen, that the $\pi^0$ region is
well described, however already in the $\eta$ region model and data disagree.
It is clear that additional sources are needed.

For a more complete view, $\Delta$ production has been added by $\pi^0$
scaling and the vector meson production by $m_T$ scaling~\cite{mt}. However,
still a factor of around 2 remains in the mass region of $0.2
GeV/c^2<M^{inv}_{ee} < 0.7 GeV/c^2$, where the $\Delta$ decay is part of the
contribution, as well as the low mass tail of the $\rho$ meson.  It has to be
clarified if this enhancement is due to a different $\Delta$ yield and/or
decay properties, or modified vector meson shapes in the dense medium. This
means that the HADES data has to be compared to advanced model
calculations\cite{hsd,rqmd,urqmd}, which on the other hand needs as much
constraints as possible extracted by elementary collisions.

\section{Summary and Outlook}
\label{sum}

In summary, results for dilepton production obtained with HADES in elementary
as well as heavy ion collisions have been presented. While the analysis of
the beam times described in this work is almost finished, the $CC$ collision
at 1AGeV is under analysis. In addition, the $pp$ run at 1.25GeV (below the
$\eta$ threshold) dedicated to $\Delta$ production has collected in spring
2006 promising statistics.  Systematic studies using elementary reactions will
be continued, which is of particular importance for the interpretation of
heavy ion inclusive mass spectra.

{\small The collaboration gratefully acknowledges the support by BMBF grants 06TM970I,
06GI146I, 06F-140, and 06DR120 (Germany), by GSI (TM-FR1,GI/ME3,OF/STR), by
grants GA CR 202/00/1668 and GA AS CR IAA1048304 (Czech Republic), by grant
KBN 1P03B05629 (Poland), by INFN (Italy), by CNRS/IN2P3 (France), by grants
MCYT FPA2000-2041-C02-02 and XUGA PGID T02PXIC20605PN (Spain), by grant 
UCY-10.3.11.12 (Cyprus), by INTAS grant 03-51-3208 and EU contract RII3-CT-2004-506078.}

%

%
%

\end{document}